\documentclass[aps,twocolumn]{revtex4}
\usepackage{amsmath,amssymb,amsthm}
\usepackage{graphicx}
\usepackage[utf8]{inputenc}
\usepackage{latexsym}
\usepackage{paralist}
\usepackage{hyperref}
\usepackage{color}
\usepackage{epsf}

\def\huga#1{\begin{gather} #1 \end{gather}}
\def\hugast#1{\begin{gather*} #1 \end{gather*}}
\def\hual#1{\begin{align} #1 \end{align}}

\def\C{{\mathbb C}}

\def\ga{\gamma}\def\del{\delta}
\def\Ga{\Gamma}\def\Om{\Omega}\def\ri{{\rm i}}\def\er{{\rm e}}
\def\al{\alpha}\def\Del{\Delta}
\def\re{{\rm Re}}
\def\sech{{\rm sech}}\def\dd{{\rm d}}
\def\pa{{\partial}}

\newcommand{\be}{\begin{equation}}
\newcommand{\ee}{\end{equation}}

\def\bmip{\begin{minipage}{\textwidth}}\def\emip{\end{minipage}}
\def\bi{\begin{itemize}}\def\ei{\end{itemize}}
\def\ben{\begin{enumerate}}\def\een{\end{enumerate}}
\def\bci{\begin{compactitem}}\def\eci{\end{compactitem}}
\def\bcen{\begin{compactenum}}\def\ecen{\end{compactenum}}
\def\bce{\begin{center}}\def\ece{\end{center}}
\newcommand{\reff}[1]{(\ref{#1})}

\newcommand{\vs}[1]{{\vspace{#1}}}

\def\eps{\varepsilon}\def\aqui{\Leftrightarrow}
\def\ra{\rightarrow}

\def\barr{\begin{array}}\def\earr{\end{array}}
\def\bpm{\begin{pmatrix}}\def\epm{\end{pmatrix}}
\def\bsm{\left(\begin{smallmatrix}}\def\esm{\end{smallmatrix}\right)}
\def\ig{\includegraphics}

\def\brem{\begin{remark}}\def\erem{\end{remark}}
\def\peps{(P$_\eps$)}\def\pnull{(P$_0$)}\def\CO{{\cal O}}
\def\beti{\tilde{\beta}}\def\psiti{\tilde{\psi}}\def\ga{\gamma}

\newtheorem{theorem}{Theorem}[section]
\newtheorem{remark}[theorem]{Remark}

\begin{document}

\title{Soliton transport in tubular networks: transmission at vertices in the shrinking limit}

\author{Hannes Uecker$^a$\thanks{hannes.uecker@uni-oldenburg.de}, Daniel Grieser$^a$, Zarif Sobirov$^b$\thanks{sobirovzar@gmail.com}, Doniyor Babajanov$^c$ and Davron Matrasulov$^c$\\ \today }

\affiliation{\vspace{3mm}
$^a$ Institut f\"ur Mathematik, Universit\"at Oldenburg, 26111 Oldenburg, Germany\\
$^b$ Tashkent Financial Institute, 60A, Amir Temur Str., 100000, Tashkent, Uzbekistan\\
$^c$ Turin Polytechnic University in Tashkent, 17 Niyazov Str., 100095,  Tashkent, Uzbekistan}

\begin{abstract}
 Soliton transport in tube-like networks is studied by solving the nonlinear Schr\"odinger equation (NLSE) on finite thickness ("fat") graphs.
The dependence of the solution and of the reflection at vertices
on the graph thickness and on the angle between its bonds is
studied and related to a special case considered in our previous
work, in the limit when the thickness of the graph goes to zero.
It is found that both the wave function and reflection coefficient
reproduce the regime of reflectionless vertex transmission studied
in our previous work.
\end{abstract}

\maketitle

\section{Introduction}\label{sec-introduction}

Particle and wave transport in branched structures is of importance
for different topics of contemporary physics such as optics, cold atom
physics, fluid dynamics and acoustics. For instance, such problems as
light propagation in optical fiber networks, BEC in network type traps
and acoustic waves in discrete structures deal with wave transport in
branched systems. In most of the practically important cases such
transport is described by linear and nonlinear Schr\"odinger equations
(NLSE) on graphs. The latter has become the topic of extensive study
during past few years \cite{Zarif, Cascaval, Banica, Adami1, Adami2, Adami3,
Adami4, Noja, Uzy4,Karim} and is still rapidly
progressing.  Such interest in the NLSE on networks is mainly caused by
possible topology-dependent tuning of soliton transport in branched
structures which is relevant to many technologically important
problems such as BEC in network type traps \cite{Leboeuf,Paul2,Oliv},
information and charge transport in DNA double helix \cite{Yomosa,Yakushevich},
light propagation in waveguide networks \cite{Burioni1} etc.

Soliton solutions of the NLSE on simplest graphs and connection formulae
are derived  in \cite{Zarif}, showing that
for certain relations between the nonlinearity
coefficients of the bonds soliton transmission
 through the graph vertex can be reflectionless (ballistic).
Dispersion relations for linear and nonlinear Schr\"odinger equations
on networks are
discussed in \cite{Banica}.
The problem of fast solitons on star graphs is
treated in \cite{Adami1} where estimates for
the transmission and  reflection coefficients are obtained in the
limit of high velocities. The problem of soliton transmission
and reflection is studied in \cite{Cascaval} by solving numerically
the stationary NLSE on graphs.
 More recent progress in the study of the NLSE on graphs can be found in
\cite{Adami2, Adami3, Adami4, Noja}.
Scattering solutions of the stationary NLSE on graphs are
obtained in \cite{Uzy4}, and analytical solutions of the
stationary NLSE on simplest graphs are derived in \cite{Karim}.

In metric graphs the bonds and vertices are one and zero
dimensional, respectively.
However, in realistic systems such as  electromagnetic waveguides and  tube-like optical fibers,
the wave (particle) motion may occur along both longitudinal and
transverse directions \cite{Markowski,Demidov,Biggs}.  Therefore it is important to study
below which (critical) thickness the transverse motions become negligible and the wave(particle) motion can be treated as one-dimensional.
In other words, studying the regime of motion when wave dynamics in such tube-like network can be considered the same as that in metric graph is of importance.

\begin{figure}[ht!]
\bce {\small \ig[width=80mm]{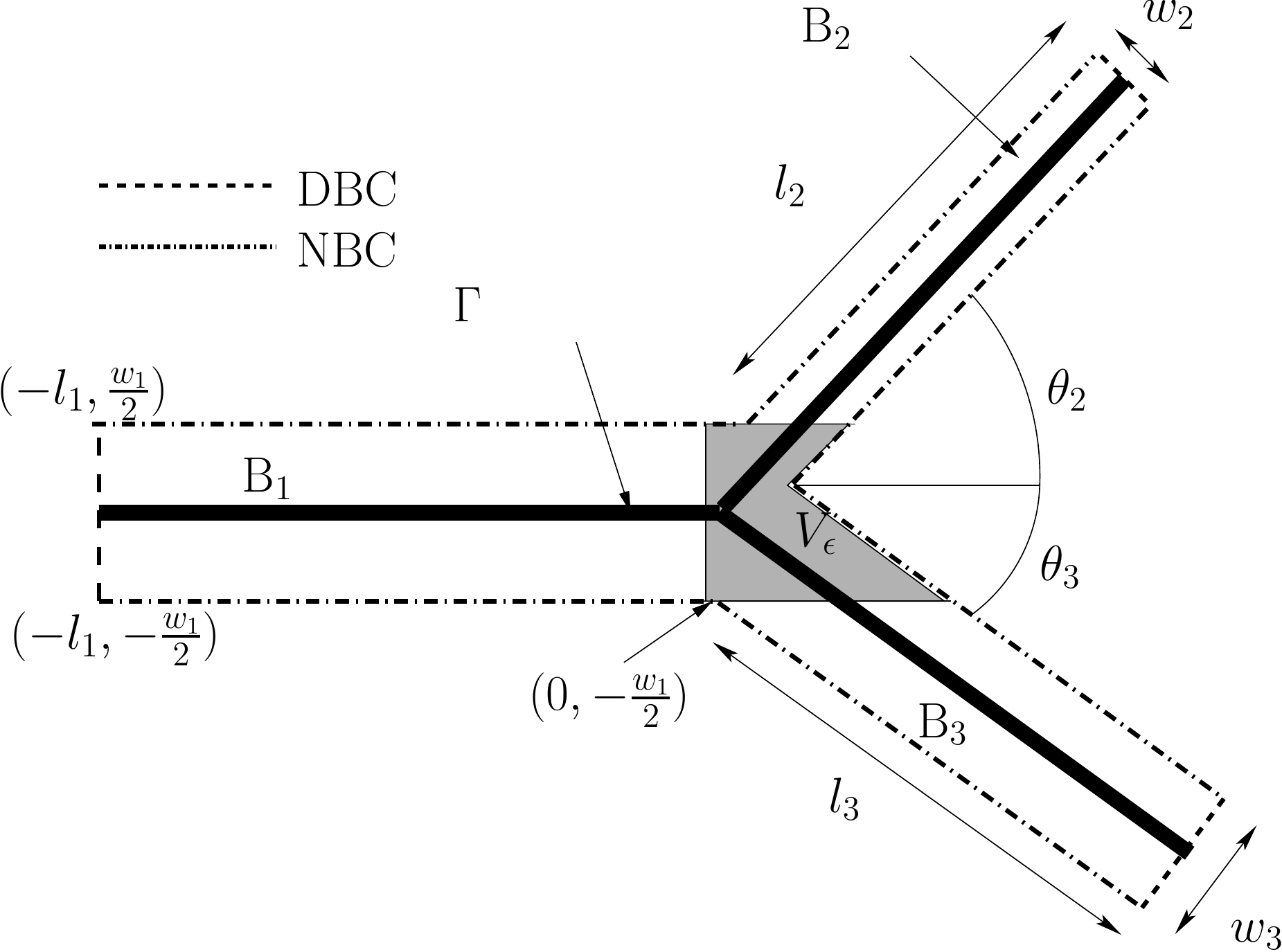} }\ece \vs{-5mm}
\caption{{\small Sketch of a metric  graph $\Ga$ and a fat graph
$\Om_\eps =V_\eps\cup B_1\cup B_2\cup B_3$, with bonds of width
$w_j$, where $w_j=\CO(\eps)$. Ideally, the lengths $l_1,l_2,l_3$
of the bonds are infinite, but for numerical simulations of the
NLSE we use finite lengths with Dirichlet boundary conditions
(DBC) at the ends, and homogeneous Neumann boundary conditions
(NBC) else. \label{pic1}}}
\end{figure}

In this paper we study the NLSE on so-called fat graphs, i.e. on two-dimensional networks having finite thickness.
The geometry will be explained in more detail below,
but see Fig.\ref{pic1} for a sketch.
In particular, we consider the same relations between the bond nonlinearity coefficients as those  in \cite{Zarif} and study the shrinking of the fat graph
into the metric graph keeping such relations. Initial conditions for the
NLSE on fat graph are taken as quasi 1D solitons.
By solving the NLSE on fat graphs we find that in the shrinking limit
such fat graphs reproduce the reflectionless regime of transport studied in \cite{Zarif}, i.e.,
the vertex transmission becomes ballistic.

The linear Schr\"odinger equation on fat graphs was the subject of
extensive study during the past decade (see,
e.g. \cite{Rubin,Kuch,PEOP1, Post, PEOP2,expo09,ex2011,expo13,cacc07,pobook,Rudenberg,Molch,MoVa10,dellCo10,kosugi2002,grieser08b}).  The first treatment of particle
transport on fat graphs dates back to Rudenberg and Scherr
\cite{Rudenberg}, who used a Green function based heuristic
approach. A pioneering study of particle transport
in fat networks comes from the paper by Mehran \cite{Mehran} on
particle scattering in microstrip bends and $Y-$
junctions, where theoretical results on reflection and transmission are
compared with experimental data.
However, the dependence of the scattering on the bond thickness and
the shrinking limit is not considered in \cite{Mehran}.

The main problem to be solved in the treatment of the Schr\"odinger equation on fat graphs is reproducing of vertex coupling rules in the shrinking limit,
i.e., when the fat graph shrinks to the metric graph.
In case of metric graphs, "gluing"  conditions, or vertex coupling rules,
are needed to ensure self-adjointness of the Schr\"odinger equation. The most important example of a vertex coupling is the Kirchhoff condition.
For fat graphs there are no such coupling rules; they only
appear in the shrinking limit, and their form
depends on specifics of the fat graph, for example
on the boundary conditions imposed at the lateral boundary.
For Neumann boundary conditions, the resulting vertex coupling is the Kirchhoff condition, as was shown in \cite{Rubin, Kuch}, who study convergence of the eigenvalue spectrum of the Schr\"odinger equation, and in a series of papers by Exner and Post \cite{PEOP1}-\cite{expo13}, who study various aspects of the Schr\"odinger equation with Neumann boundary conditions (including transport, resonances and magnetic field effects).
The vertex couplings obtained in the shrinking limit of the Schr\"odinger equation on the fat graph with Dirichlet and other boundary conditions were obtained in
\cite{Molch, grieser08b}.
Recent studies of the linear Schr\"odinger equation on fat graphs focused on the inverse problem of finding a suitable fat graph problem which reproduces a given coupling rule in the shrinking limit \cite{cacc07}. Further references on linear Schrodinger equation on fat graphs are
 \cite{Exner1,Exner2,Kost,Uzy1,Uzy2,Uzy3,dellCo10,MoVa10,ex2011,expo13}, and the reviews
\cite{grieser08,pobook}. All the above results have been  limited  to linear and stationary cases, and spectral results.
Related problems also
have a long history in (nonlinear) PDEs, see \cite{raugel95} and
the references therein, where however the focus is on dissipative
systems, and on damped wave equations.

The case of the NLSE on fat graphs is much more
complicated than the linear case. Therefore one may expect that the
treatment of the NLSE  with the same success as for the
linear problem is not possible.
To our knowledge, the only work dealing with nonlinearities on
fat graphs is by Kosugi
\cite{kosugi2002}, who considers semilinear elliptic
problems and  shows $L^\infty$ convergence of
solutions towards solutions of the metric graph problem. However,
for problems such as soliton transport, scattering and interaction with external
potentials which are described by time-dependent evolution
equations on fat graphs, we have to rely to a large extent on numerics.

In this paper, using the numerical solution of the
NLSE on fat graph we explore the dependence of soliton transmission and
reflection at the fat graph vertex on the bond thickness and the
angle between the bonds. It is organized as follows. In the next
section we give detailed formulations of the problems both for fat
and metric graphs. Section III presents numerical (soliton)
solutions of the NLSE on fat graphs, and analysis of the soliton reflection
at the graph vertex in the shrinking limit, including the dependence of
reflection coefficient on the angle between the graph bonds.
Section IV gives conclusions, while the Appendix contains some details
of the numerics.

\section{The NLSE on metric and fat graphs}

Consider the nonlinear Schr\"odinger equation
\huga{
\pa_t\psi_k=\ri(\psi_k''+\beta_k|\psi_k|^2\psi_k),\quad k=1,2,3,
\label{qgnls} } on a metric star graph $\Ga$ with $3$ edges $\Ga_{k}$,
and nonlinearity coefficients $\beta_k>0$. The graph is assumed to have semi-infinite bonds
 $\Ga_1=(-\infty,0)$, $\Ga_{2,3}=(0,\infty)$,
but the main part of our analysis will be numerical, for which we
assume finite lengths $l_k$ of bonds, with coordinates $\xi_1\in
(-l_1,0)$, $\xi_{2,3}\in (0,l_{2,3})$, and
 homogeneous  Dirichlet boundary conditions  at
$\xi_1=-l_1, \xi_{2,3}=l_{2,3}$. Furthermore, we assume that the
solutions, $\psi_k=\psi_k(t,\xi_k)\in\C$  obey the
vertex (at $\xi_k=0$) conditions
\huga{\al_1\psi_1=\al_2\psi_2=\al_3\psi_3,\quad \frac 1 {\al}_1\psi_1'
=\frac 1 {\al_2}\psi_2'+\frac 1 {\al_3}\psi_3',
\label{vc}}
with parameters $\al_k$, where it is
understood that $\psi_1'$ ($\psi_{2,3}'$) denote the derivatives
from the left (right). In the following  we call Eqs.\reff{qgnls}
and \reff{vc} problem \pnull.

\begin{figure}[t!]
\begin{tabular}{ll}
\ig[width=40mm,height=50mm]{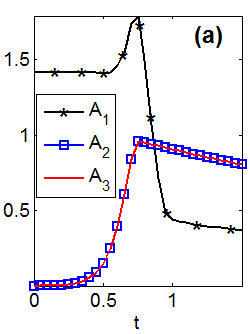}&
\ig[width=40mm,height=50mm]{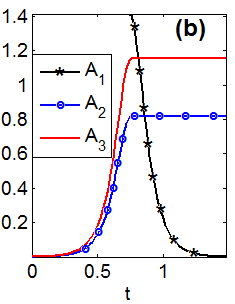}
\end{tabular}
\caption{{\small (Color online) Amplitudes
$A_k=\max_{x\in\Ga_k}|\psi_k(t,x)|$ for \reff{qgnls} on the metric
graph $\Gamma$ with bond lengths $15$. Initial soliton of the form
\reff{psia} with $\eta,c=1,10$ and $\xi_0=-7.5$, see also
\reff{gaic}. (a) Kirchhoff case, $\alpha=(1,1,1)$,
$\beta=(1,1,1)$; (b) ballistic case,  $\alpha=(1,1.73,1.22)$,
$\beta=(1,3,1.5)$. In (a), the blue line is hidden by the red
line.\label{f1}}}
\end{figure}

Soliton solutions of the problem
\pnull\ that propagate without reflection (i.e., ballistically)
were obtained analytically
in \cite{Zarif} for the special case when the nonlinearity coefficients satisfy
the relation
\be \frac{1}{\beta_1} =\frac{1}{\beta_2}
+\frac{1}{\beta_3}. \label{ballist}
\ee
These solutions have, after
properly identifying $\xi$ with $\xi_k$ on $\Gamma_k$ the form
\huga{\label{psia} \psi_{k}(t,\xi)=\frac
{\sqrt{2}}{\sqrt{\beta_k}}\eta\sech(\eta(\xi{-}\xi_0{-}ct))
\er^{-\ri(2c\xi-(c^2-4\eta^2)t)/4},
}
with free parameters
amplitude $\eta>0$, speed $c$ (wavenumber $c/2$), and reference
position $\xi_0$. Fig.\ref{f1} presents amplitudes, $A_k=\max_{x\in\Ga_k}|\psi_k(t,x)|$ for  Kirchhoff boundary conditions ($\alpha_1=\alpha_2=\alpha_3=1$) and
for the boundary conditions given by Eq.\reff{vc}.
The vertex boundary conditions given by \reff{vc} are
one possibility to make the linear part of \reff{qgnls}
skew-adjoint. The problem \pnull\ conserves the
norm $N$ and the Hamiltonian $H$ given by \hual{
&N{=}\sqrt{N_1^2+N_2^2+N_3^2},\quad
N_k^2(t){=}\int_{\Ga_k}|\psi_k(t,x)|^2\dd \xi, \\
&H=H_1+H_2+H_3, \nonumber\\
&\quad H_k(t)=\int_{\Ga_k} |\pa_\xi\psi_k(t,\xi)|^2
-\frac{\beta_k} 2 |\psi_k(t,x)|^4 \dd \xi. } It is a question of
normalization to set \huga{\label{nsc1} \al_1=\beta_1=1, } which
leaves 4 parameters for \pnull, and, of course, the choice of the
initial conditions.

Our goal is to compare exact and numerical
solutions $(\psi_1,\psi_2,\psi_3)$
of \pnull\ with the numerical solutions $\phi=\phi(t,x)$
of an associated NLSE on a fat graph presented in Fig.~\ref{pic1}, i.e.,
\huga{
\pa_t\phi=\ri(\Delta\phi+\beti(x)|\phi|^2\phi),\label{tbmnls}
}
where $\Delta=\pa_{x_1}^2+\pa_{x_2}^2,\ x=(x_1,x_2)\in\Om_\eps$, and
 $\Om_\eps=V_\eps\cup B_{1,\eps}\cup B_{2,\eps}\cup
B_{3,\eps}$ consists of a ``vertex--region'' $V_\eps$ of diameter
$\CO(\eps)$, and $\CO(\eps)$-tubes $B_k$ around $\Ga_k$, see
Fig.~\ref{pic1}. In the following Eq.\reff{tbmnls} will be called
the problem \peps.
We also use the notation $\phi_k$ for
$\phi|_{B_k}$.

It is clear that different versions of $\Om_\eps$ are possible.
Here we choose to give the following 5 parameters to
$\Om_\eps$ not a priori present in Eq.\reff{qgnls}:
\bcen
\item the angles $\theta_2,\theta_3$ between the bonds $B_2$ and $B_3$ and
the $x_1$--axis,
\item the widths $w_1,w_2,w_3$ of the different bonds.
\ecen
In the numerical calculations we impose homogeneous Dirichlet
boundary conditions (DBC) for both, \pnull\ and \peps,
at the ``ends'' of bonds, and for \peps\
homogeneous Neumann boundary conditions (NBC) $\pa_n \psi=0$
everywhere else. As our simulations will run on time--scales where
the solitons will be well separated from the ends of the bonds, we
could as well pose NBC there. Also note that strictly
speaking \reff{psia} is not a solution over the finite graph, but
it is exponentially small at the ends of the bonds.

We take $\beti(x)$ constant on bond $k$ and with suitable jumps
near $0$.
Furthermore, we set
\huga{\text{$\eps:=w_1$,\
$w_2=\del_2\eps$ and $w_3=\del_3\eps$}
}
and write $\Om_\eps$ for
fixed $\del_k,\theta_k$, $k=2,3.$ For
definiteness we choose \huga{\label{bkdef}
B_1=\Om_{\eps}\cap\{x_1<0\}, \quad
B_2=\Om_{\eps}\cap\{x_2>w_1/2\}, \nonumber\\
 \quad
B_3=\Om_{\eps}\cap\{x_2<-w_1/2\}, } and thus
$V_\eps=\Om_\eps\setminus (B_1\cup B_2\cup B_3)$. Motivated by
$\frac 1 \eps \int_{\Om_\eps} 1 \dd x\ra l_1+\del_2l_2+\del_3l_3$
as $\eps\ra 0$, corresponding to $N$ on $\Gamma$ we define the
scaled norms
\huga{
 N_\eps(t)=
\left(\frac 1 \eps \int_{\Om_\eps} |\phi(t,x)|^2\dd
x\right)^{1/2},\\
\text{ and } N_{k,\eps}(t):=\left(\frac 1 \eps \int_{B_k}
|\phi(t,x)|^2\dd x\right)^{1/2}. \label{nepsdef}
} Then $N_\eps$ is conserved for
\reff{tbmnls}, and the $N_{k,\eps}$ indicate how much ``mass'' is
in the different bonds.

For the linear problem it is known, \cite{expo09}, that under the
scaling \huga{\label{sc1} \frac  {w_1}{w_k}=\al_k^2,\text{ i.e.
}\del_k=\frac 1 {\al_k^2}, \text{ and }
\psi_k=\frac{1}{\al_k}\phi_k|_{\Ga_k}, } the vertex conditions
\reff{vc} appear in the limit $\eps\ra 0$. Then, at least
formally, we can expect \pnull\ as a ``limit'' of \peps\ if
\huga{\beti|_{B_k}=w_k \beta_k= \al_k^{-2}\beta_k.\label{beti} }
If $\al_2\ne 1$ (or $\al_3\ne 1$), then the boundary conditions \reff{vc}
gives jumps from $\psi_1$ to $\psi_2$ (resp.~$\psi_3$) at the
vertex. This, however, is merely a question of scaling. For
instance, setting $\psiti_k=\al_k\psi_k$ (cf.~\reff{sc1}), we
obtain \hual{
\pa_t\psiti_k&=\ri(\psiti_k''+\ga_k|\psiti_k|^2\psiti_k),\quad
\psiti_1=\psiti_2=\psiti_3, \nonumber\\
&\quad \psiti'_1=\frac 1 {\al_2^2}\psiti_2' +\frac 1
{\al_3^2}\psiti_3',  \text{ at $x=0$}, \label{rsc2} } i.e.,
continuity at the vertex, where $\ga_k=\beta_k\al_k^{-2}$, as in
\reff{beti}. The scaling given by Eqs.\reff{qgnls},\reff{vc} is more custom
\cite{Zarif,expo09} than \reff{rsc2}, and therefore we stick to
\reff{qgnls},\reff{vc} as the ``limit problem''. Note that the angles
$\theta_{1,2}$ of the fat graph do not appear
in \pnull.

We expect that for $\eps\ra 0$ solutions $\phi_k$ of \peps\ behave like
$\frac 1 {\al_k} \psi_k$ with  $\psi_k$ being the solutions of \pnull, i.e.,
 are constant in transverse direction on each bond
$B_k$, with width $w_k=\del_k\eps$. Therefore, from Eqs.~\reff{nepsdef} and
\reff{sc1} we expect
\huga{ N_{k,\eps}^2(t)=\frac 1 \eps
\int_{B_k} |\phi_k(t,x)|^2\dd x \approx
\del_k\int_{\Ga_k}|\phi_k|_{\Ga_k}|^2\dd \xi_k \nonumber\\
\approx \del_k
\int_{\Ga_k}|\al_k|^2|\psi_k|^2\dd \xi_k =N_k^2(t),
}
In the numerical calculations, in addition to $N_{k,\eps}$
we explore the following functions (dropping the dependence on parameters
$\eps,\del_{2,3},\theta_{2,3}, c$ and $\eta$):
\hual{
A_{k}(t)=&\frac 1 {\al_k} \max_{x\in B_k}|\phi_k(t,x)|
\quad \text{(scaled amplitude),}\\
m_{k}(t)=&\max_{x\in B_k}\bigl| |\psiti_k(t,x)|
-\frac{1}{\al_k}|\phi_k(t,x)|\bigr| \label{mkdef} \\
\text{ (maximal}&\quad\text{amplitude distance between \peps\ and \pnull).}
\nonumber
}
Here $\psiti_k$ is the extension of $\psi_k$ to $B_k$, constant in
transverse direction, and for $\psi_k$ we either use the explicit
formula \reff{psia} if \reff{ballist} holds, or  numerics for \pnull\
if not. Note that \reff{mkdef} ignores phase differences
between $\psiti_k$ and $\phi_k$, as these are less important from the viewpoint of applications.

\section{Soliton transport in fat graphs}
The main practically important problem in the context of wave
propagation in branched systems is energy and information
transport via solitary waves. The dependence of the soliton
dynamics on the topology of a network makes  such systems
attractive from the viewpoint of tunable transport in low
dimensional optical, thermal and electronic devices. Therefore,
the treatment of the problems \pnull\ and \peps\ from the
viewpoint of vertex soliton transmission is of importance. Our
main purpose is to compare propagation of solitons in $\Om_\eps$
with that in $\Gamma$ in particular we are interested to study the
``lift'' the earlier results \cite{Zarif} from $\Ga$ to
$\Om_\eps$. Transition from two- to one-dimensional wave motion in
the shrinking limit is of special importance for this analysis.

In a typical simulation, for \pnull\ we use soliton-type initial condition given as
\huga{\label{gaic}
\psi_1(0,\xi_1)=\sqrt{2}\eta\,\sech(\eta(\xi_1-x_0)) \er^{-\ri
c\xi_1/2}, \quad \psi_{2,3}(0,\cdot)\equiv 0
} where $x_0$ and
$\eta$ are chosen in such a way that $\psi_1(0,0)$ is very close
to $0$. Similarly, for \peps\ we choose
\huga{\label{epsic}
\phi(0,x){=}\left\{\barr{ll}\sqrt{2}\,\eta\sech(\eta(x_1{-}x_0))
\er^{-\ri cx_1/2} &\ x_1<0,\\
0&\ \text{else}, \earr\right. } i.e., we extend the initial
conditions \reff{gaic} trivially in transverse direction. We then
run both, \pnull\ and \peps\ until some final time $t_1$ such that
the solitons launched by \reff{gaic} and \reff{epsic},
respectively, have interacted with the vertex, and have been
reflected or transmitted sufficiently far into the bonds (see the
appendix for the numerical methods used). Our main solution
diagnostics will be the time dependent norms $N_k(t),
N_{k,\eps}(t)$, the amplitudes $A_k(t), A_{k,\eps}(t)$, the
distances $m_k(t)$, and the reflection coefficients defined below.

\begin{figure}[t!]
\resizebox{1\columnwidth}{!}{%
\begin{tabular}{llll}
\ig[width=36mm]{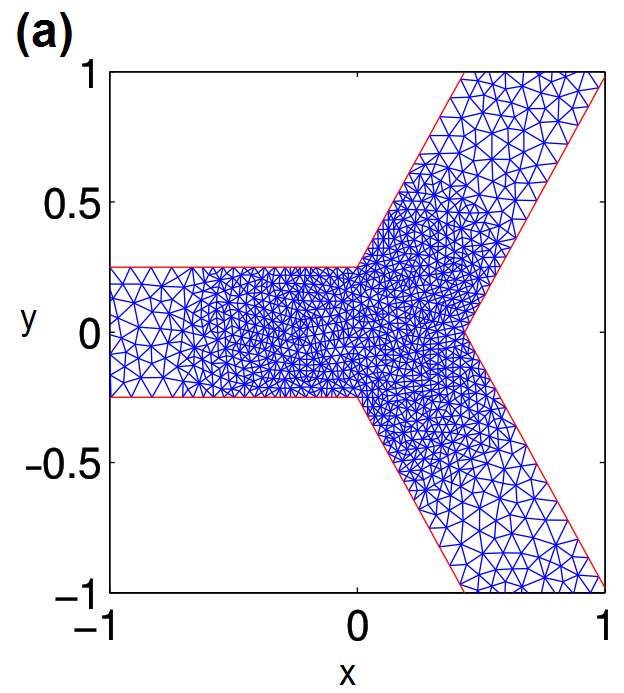}&
\ig[width=36mm]{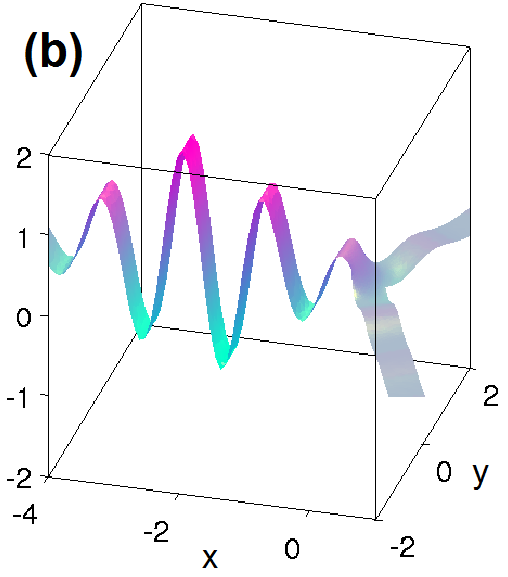}\\
\ig[width=36mm]{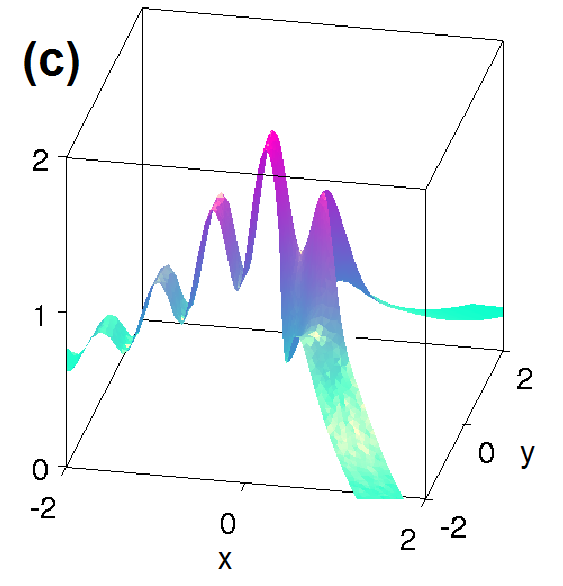}&
\ig[width=36mm]{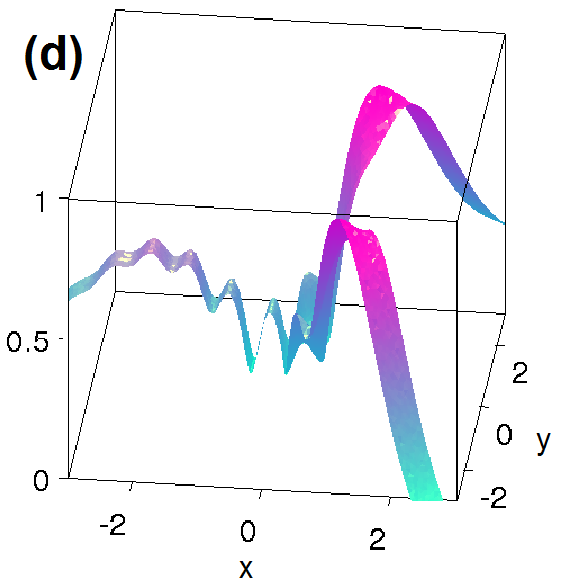}\\
\end{tabular}
 }
 \caption{{(Color online) Numerical solution of \peps\
for $\del_{2,3}=1$ and $\eps=0.5$, i.e.~$w=(0.5,0.5,0.5)$;
$\beti\equiv 1$, $l=(15,15,15)$, $\theta=(\pi/3,\pi/3)$. Initial
condition \reff{epsic} with $x_0=-l/2$ and $\eta,c=1,10$. (a)
Geometry and mesh near the vertex. (b) $\re\psi(0.5,\cdot)$ real
part of incoming soliton at $t=0.5$; (c),(d) $|\psi(\cdot,x)|$
during and after transmission/reflection trough/at the vertex.
\label{f2c}}}
\end{figure}

\begin{figure}[t!]
\resizebox{0.9\columnwidth}{!}{%
\ig[height=40mm]{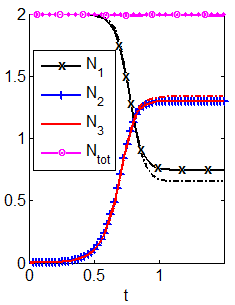} \ig[height=40mm]{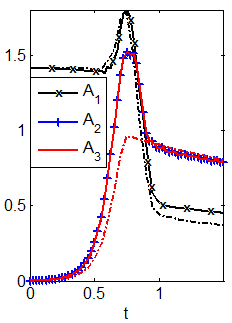} }
\caption{(Color online) Norms and  amplitudes corresponding to the
solutions presented in Fig.\ref{f2c}. Dashed lines present
 respective quantities from
\pnull; the blue lines are all hidden by the red lines. \label{f2e}}
\end{figure}

For definiteness, we consider $\Gamma_1$ as the ``incoming'' bond
and $\Gamma_{2,3}$ as ``outgoing'' ones. In Fig.~\ref{f2c}
solutions of the problem \peps\ for the Kirchhoff boundary
conditions are presented for the case of a ``relatively fat''
graph ($\eps=0.5$), while Fig.~\ref{f2e} shows the plots of the
corresponding norms $N_k$ and amplitudes $A_k$ for the simulation
for \peps\ in Fig.~\ref{f2c} (Kirchhoff case), together with the
respective quantities for \pnull. At this relatively large
$\eps=0.5$ there is a significant difference between \peps\ and
\pnull.

\begin{figure}[t!]
\resizebox{0.9\columnwidth}{!}{%
\begin{tabular}{ll}
\ig[width=40mm]{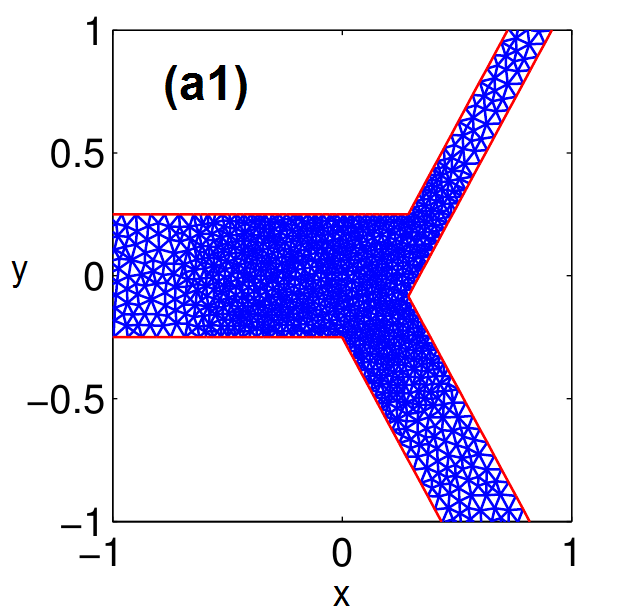}&\ig[width=40mm]{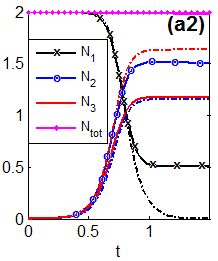}\\
\ig[width=40mm]{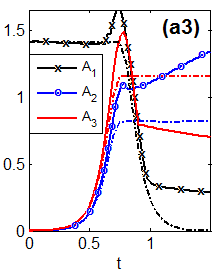}&\ig[width=40mm]{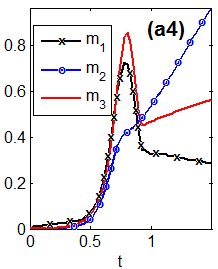}\\
\end{tabular}
}
\caption{(Color online) Norms and amplitudes for
 fat and metric graphs with
$1/\al_2^2+1/\al_3^2=1$ and $\beta_k=\al_k^2$, hence $\beti=1$,
and plots of the amplitude distances $m_{k,\eps}$,
cf.~\reff{mkdef}. Here $\theta_1=\theta_2=\pi/3$, $\delta_2=1/3,
\delta_3=2/3$, and $\eps=0.5$, hence $w=(0.5,0.33,0.17)$.(see
Fig.~\ref{f21}). In (a1) we also plot the geometry and mesh near
the vertex. For the lengths of the bonds we again have
$l_1=l_2=l_3=15$. In (a2),(a3) the full lines are $N_{\eps,k}$ and
$\frac 1 {\al_k} A_{\eps,k}$, respectively, and the dashed lines
are $N_k$ and $A_k$, cf.~Fig.~\ref{f1}(b), and similarly in
(b1),(b2) and (c1),(c2). \label{f2}}
\end{figure}

\begin{figure}[t!]
\resizebox{0.85\columnwidth}{!}{%
\begin{tabular}{ll}
\ig[width=30mm]{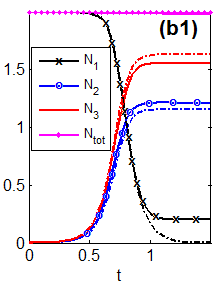}&\ig[width=30mm]{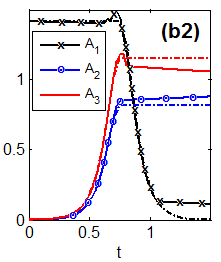}\\
\ig[width=30mm]{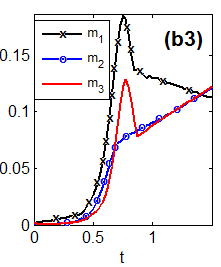}&\ig[width=30mm]{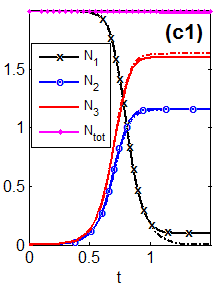}\\
\ig[width=30mm]{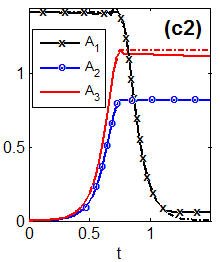}&\ig[width=30mm]{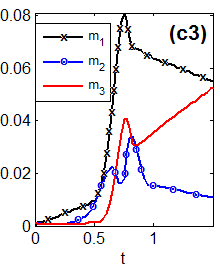}
\end{tabular}
} \caption{(Color online) Continuation of Fig.5, $\eps=0.2$ in
(b) and $\eps=0.1$ in (c) \label{f21}}
\end{figure}

In the following we focus on soliton  reflection and transmission
in the shrinking limit, $\varepsilon \to 0$ for the "ballistic"
boundary conditions given by \reff{ballist} on \pnull. In
Fig.\ref{f2} and \ref{f21} we plot  the diagnostics defined above
for different $\eps$ on an otherwise fixed graph fulfilling
\reff{ballist}, i.e., for the ballistic case. As $\eps\ra 0$, the
amplitudes and masses in the different bonds get close to the
metric graph case, and also the (numerical) wave functions as a
whole converge to the ones on the metric graph, with one small
qualification: While the main mismatches between \peps\ and
\pnull\ result from reflection and position shifts of the incoming
soliton during interaction with the vertex around $t=7.5$, already
for $0<t<5$, i.e.,  before interaction of the soliton with the
vertex, there is a small linear growth of $m_{1,\eps}$, i.e., of
the amplitude mismatch in the incoming bond. This is not a
property of the fat graph itself, but related to the fact that it
is difficult to accurately resolve the speed of the soliton
numerically. In other words, for small $\eps$, a significant part
of mismatch between our (numerical) fat graph solution $\phi$ and
the (analytical) metric graph solution $(\psi_1,\psi_2,\psi_3)$
from \cite{Zarif} is not due to the behaviour at the vertex, but
due to an error in (numerical) soliton speed, which results in a
position mismatch growing in time (see the Appendix for further
discussion). However, noting the different scales in panels
(a4),(b3) and (c3) strongly indicates the convergence of the
\peps\ wave function to the \pnull\ wave function in $L^\infty$
(modulo phases), uniformly on bounded time intervals.

From the  viewpoint of practical applications, probably the most
important question is how much of an incoming soliton is reflected
resp.~transmitted in the vertex region of a fat graph. To display
this in a concise way, for \peps\ we define the reflection and
transmission coefficients \hual{
r_{k,\eps}^A&:=A_{k,\eps}(t_1)/A_{1,\eps}(0) \quad\text{(amplitude
reflection)},\label{rAdef}\\
r_{k,\eps}^N&:=N_{k,\eps}(t_1)/N_{1,\eps}(0)\quad\text{(mass
reflection)},\label{rNdef} } where again we dropped the dependence
on parameters $w_{2,3},\beta_{2,3}, c$ and $\eta$ here, but will
plot $r_{k,\eps}^A, r_{k,\eps}^N$ as functions of some parameters
below. Thus, e.g., $r_{1,\eps}^A=0$ (and thus also
$r_{1,\eps}^N=0$) means zero reflection of an incoming soliton at
the vertex, while, e.g., $r_{2,\eps}^N=1$ means that all of the
``mass'' was transmitted to bond two. These extreme cases of
course do not occur, but the goal is, e.g., to tune
$r_{k,\eps}^{N,A}$. The corresponding quantities for \pnull\ are
defined as \huga{ r_{k}^A:=A_{k}(t_1)/A_{1}(0),\quad
r_{k}^N:=N_{k}(t_1)/N_{1}(0), \label{rNdef0} } and the
transmission formula \reff{ballist} means, that $r_1^{A,N}\ra 0$
in the limit of infinite bonds and of $t_1\ra \infty$.

\begin{figure}[t!]
\begin{tabular}{ll}
\ig[width=40mm,height=40mm]{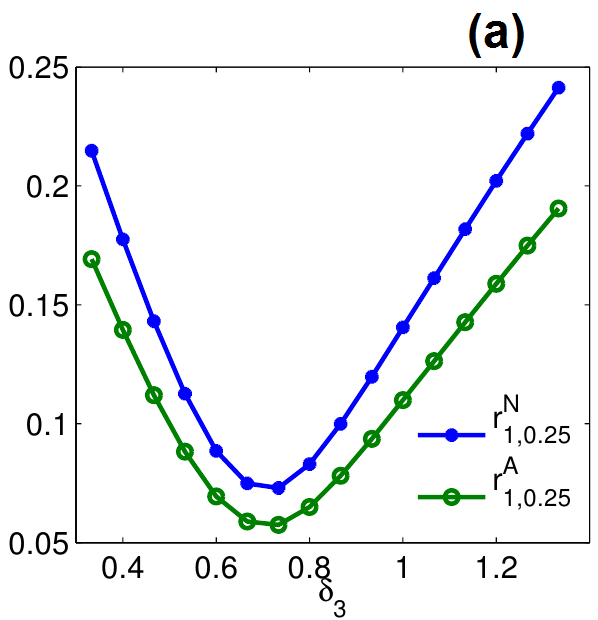}&
\ig[width=40mm,height=40mm]{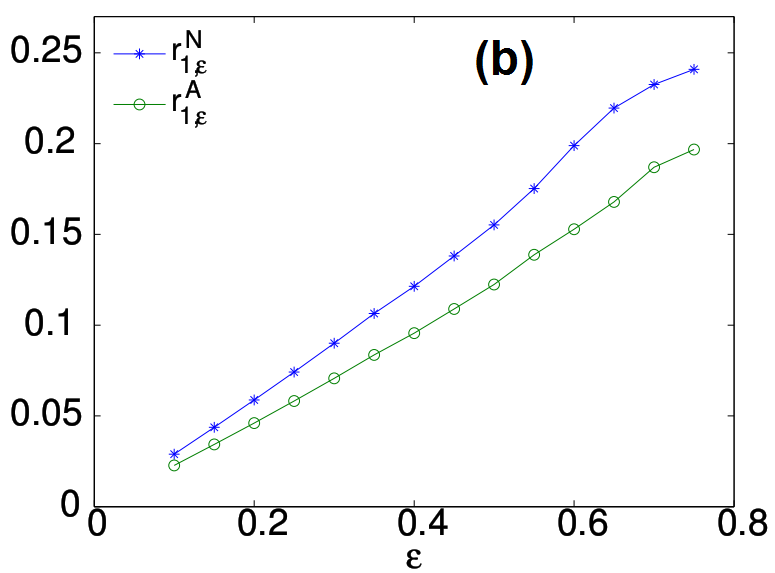}
\\
\end{tabular}
\resizebox{0.8\columnwidth}{!}{%
\begin{tabular}{l}
\ig[width=60mm]{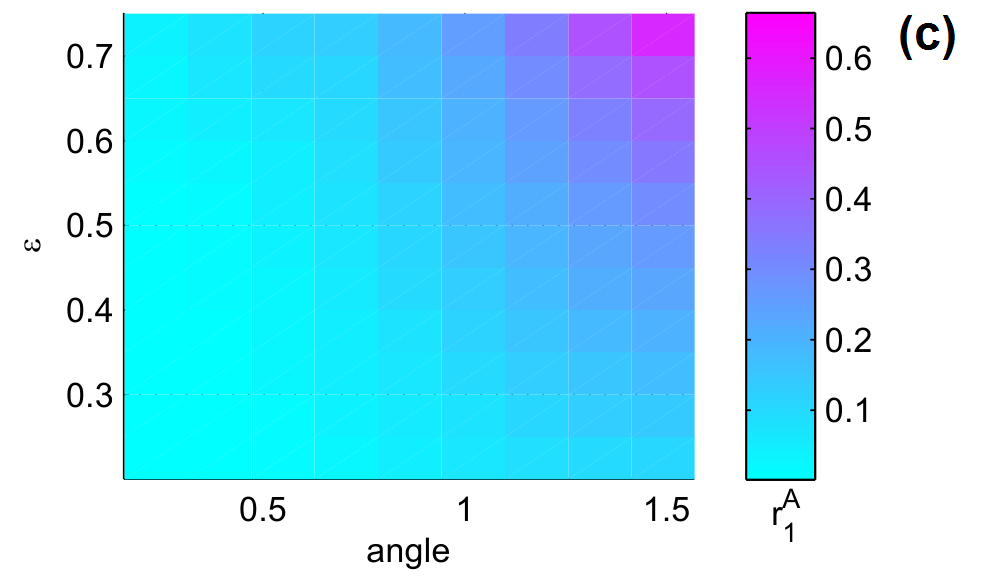}
\end{tabular}
} \caption{(Color online) (a)$r_{k,\eps}^N$ and $r_{k,\eps}^A$ as
functions of the (relative) thickness $\del_3$ of the third bond,
with fixed $\varepsilon=0.25$,$\theta_{1,2}=\pi/4$,$\del_1=1,
\del_2=1/3$,$l=(15,15,15)$, and $(\eta,c)=(1,10)$. (b)
$r_{k,\eps}^N$ and $r_{k,\eps}^A$ as functions of $\eps$, with
$\del_3=2/3$ (ballistic case) and the remaining parameters fixed
as in (a). (c) $r_{1,\eps}^A$ as a function of $\eps$ and angles
$\theta:=\theta_1=\theta_2$, remaining parameters as in (b).
\label{db1}}
\end{figure}

Fig.\ref{db1} presents plots of the reflections as a function of
different parameters such as bond thickness, $\varepsilon$, angle
between the bonds, $\theta$ and coefficient, $\delta_3$. Of
course, the ballistic regime is an idealization, and in order to
quantify the reflections when we perturb it, in Fig.~\ref{db1}(a)
we study the dependence of $r_{k,\eps}^N$ and $r_{k,\eps}^A$ on
the mismatch $\frac 1 {\beta_1}-\frac 1 {\beta_2} -\frac 1
\beta_3$. As we have chosen $\beta_k=w_k^2=1/\del_k$, $k=2,3$, in
(a) we fix $\del_2=1/3$ and let $\del_3$ vary between $1/3$ and
$4/3$, such that again $\del_3=2/3$ corresponds to the minimal
reflection case.
Clearly, the minima of $r_{k,\eps}^N$ and $r_{k,\eps}^A$ are
attained close to $\del_3=2/3$, and these functions are somewhat
steep. Similar graphs were obtained in other geometries (e.g.,
angles), and thus in applications it appears desireable to move as
close as possible to the ballistic case by, e.g., fine tuning the
widths of the bonds. In Fig.~\ref{db1}(b) the vertex reflection
coefficients (both for norm and amplitude) are plotted as
functions of the graph thickness $\varepsilon$ for the case
described by \reff{ballist}. The limit $\eps\ra 0$ again shows a
rather smooth transition from the ``scattering'' to ballistic
regime.Figure \ref{db1}(c) presents the dependence of
$r_{1,\eps}^A$ on the graph thickness and the angles
$\theta=\theta_1=\theta_2$. Even though the angles do not appear
in the $\eps\ra 0$ limit \pnull, at finite $\eps$ they of course
play an important role. Ballistic transport through the vertex
occurs in the shrinking limit as well as in the limit of small
angles.

\begin{figure}[t!]
\begin{tabular}{ll}
\ig[width=40mm,height=40mm]{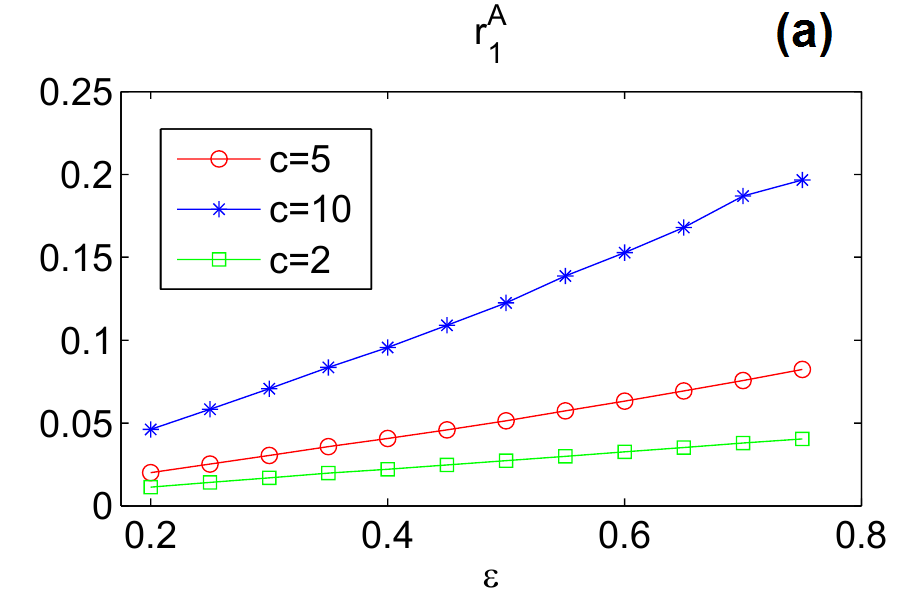}&
\ig[width=40mm,height=40mm]{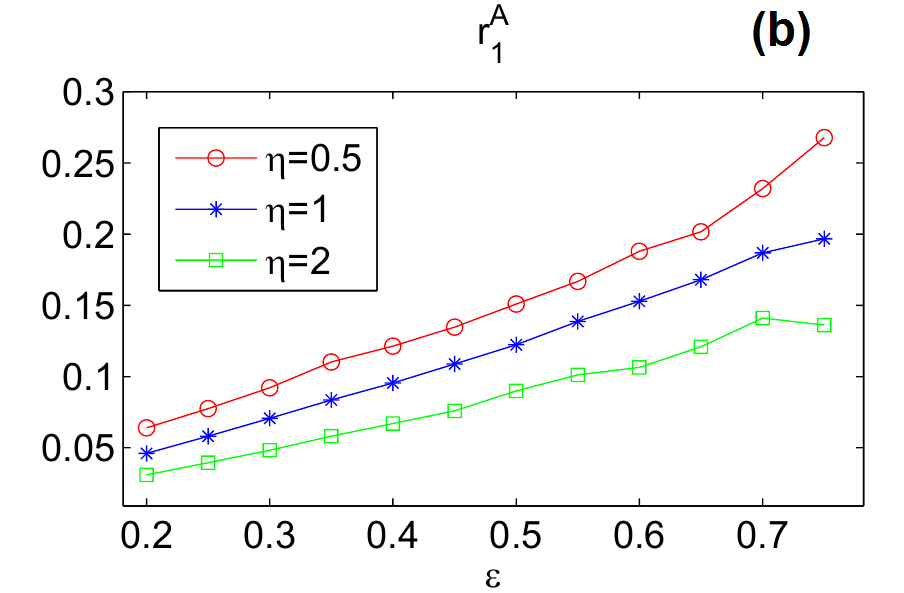}
\end{tabular}
\caption{{\small (Color online) Reflection coefficients as
functions of thickness $\varepsilon$, with fixed
$\theta_{1,2}=\pi/4$, $l=(15,15,15)$ for differrent values of (a)
speeds (with fixed $\eta=1$) and (b) amplitudes (with fixed
$c=10$) of soliton in the ballistic case.\label{fig8}}}
\end{figure}

Besides the equal angle case $\theta_2=\theta_3$, we checked a
variety of other configurations with $\theta_2<\theta_3$,
for various $\theta_{1,2}$ between $\pi/20$ and $\pi/2$.
The results remain qualitatively similar to Figs.~\ref{f2}--\ref{db1}, i.e.,
in the ballistic case the reflection coefficients vanish as $\eps\ra 0$, and
as above the
\peps\ wave functions converge to the $\theta_{2,3}$ independent
wave functions $(\psi_1,\al_2\psi_2,\al_3\psi_3)$ of \pnull.
As the convergence
for $\eps\ra 0$ is clearly linear, an interesting question is
how to choose a first order in $\eps$ correction of the
fat graph geometry or NLSE coefficients that minimizes
$r_{1,\eps}^{N,A}$ also for finite $\eps>0$.

An important issue for particle and wave  transport in fat graphs
is the dependence of the scattering on initial soliton  velocity
and amplitude. In Fig.8 reflection coefficients are plotted as
functions of bond thickness $\varepsilon$ for different initial
velocities (a) and amplitudes (b). The dependence of reflection on
initial data is significant for fat graphs, with, e.g., less
reflection for slower and longer waves, as should be expected.
However, in the shrinking limit the reflections vanish in all
cases considered.

Finally, although in Figs.~\ref{f2}--\ref{db1} we mainly focused on the ballistic
case $\delta_2+\delta_3=1$,  for other values of
$\delta_2,\delta_3$, as for instance $\delta_2=\delta_3=1$ in Fig.~\ref{f2e},
as $\eps\ra 0$ we have the same kind of convergence
of $(N_{k,\eps}, \frac 1 {\al_k}A_{k,\eps}, r_{k,\eps}^{N,A}, m_{k,\eps})$ to
$(N_{k}, A_{k}, r_{k}^{N,A}, 0)$ as above, and
altogether of $\phi$ to $(\psi_1,\al_2\psi_2,\al_3\psi_3)$, i.e.,
of \peps\ to \pnull.

\section{Conclusions}
We studied soliton transport in tube like networks modeled
by the time-dependent NLSE on fat graphs, i.e.~graphs with finite bond
thickness.
We numerically solved the NLSE on fat graphs for different values of
thickness, and 
studied behavior of solutions and vertex reflection coefficients
in the shrinking limit. It is found that in the shrinking limit
solutions of the NLSE on fat graphs converge to those on the
associated metric graphs, and hence the conditions \reff{ballist}
for reflectionless transport also work on fat graphs with small
$\eps$. The dependence of the vertex reflection coefficient on the
bond thickness and on the angle between the bonds of fat graph is
also studied.

At this point it is not clear in which norms we can expect or
analytically show convergence of solutions of \peps\ to solutions
of \pnull, as $\eps\ra 0$.  First, following \cite{kosugi2002}
this will be discussed for the stationary case, including some
potentials at the vertex in order to have nontrivial stationary
solutions for the fat graph and the metric graph,
cf.~\cite{Adami2, Karim}. An important point in the study of
wave(particle) dynamics in fat graphs is the definition of the fat
graph thickness at which one can neglect transverse motion and
consider the system as one-dimensional. The above treatment allows
us to define such a regime. However, the transition from two- to
one dimensional motion is rather smooth and there is no critical
value of the bond thickness at which a "jump" from the fat to the
metric graph occurs.  In any case, we believe that our numerical
results should be considered as a first step in the way for the
study of particle and wave transport described by nonlinear
evolution equations on fat graphs. In addition, can be useful for
further analytical studies of the NLSE on such graphs.

\section*{Acknowledgement} We thanks Pavel Exner and Riccardo Adami for discussions and useful comments. This work is supported by a grant of
the Volkswagen Foundation.

\section*{Appendix: Details of the numerical approach}
We discretize \pnull\ by second order spatial finite differences (FD)
and denote $u_j{=}u_j(t){=}\psi_{1}(t,\xi_{1,j})$, $\xi_j{=}-l_1+j\del$,
$v_j=v_j(t)=\psi_{2}(t,\xi_{2,j}),
w_j=w_j(t)=\psi_{3}(t,\xi_{3,j})$, $j{=}1,\ldots,n-1$, such that,
e.g., $u_j''=\frac 1 {\del^2}(u_{j-1}-2u_j+u_{j+1})$. Moreover, we
set
\hugast{
\text{$u_0=\psi_1(-l_1)=0$, $v_n=\psi_2(l_2)=0$,
$w_n=\psi_3(l_3)=0$,}\\
\text{ and $u_n=\psi_1(0)$, $v_0=\psi_2(0)$, and $w_0=\psi_3(0)$.}
}
The vertex conditions then are
$u_n{=}\al_2
v_0{=}\al_3w_0$, $u_n'=\frac 1 {\al_2} v_0'+\frac 1 {\al_3} w_0'.$
Using one-sided FD for $u_n',v_0'$ and $w_0'$ we have \hual{
u_n'&=\frac 1 \del(u_n-u_{n-1})=\frac 1 \del\left(\frac 1
{\al_2}(v_1-v_0)
+\frac 1 {\al_3}(w_1-w_0)\right),\notag \\
&\aqui u_n(1+\frac 1 {\al_2^2}+\frac 1 {\al_3^2})= u_{n-1}+\frac 1
{\al_2} v_1+\frac 1 {\al_3}w_1.\label{undef} } which expresses
$u_n$ and hence $v_0,w_0$ in terms of $u_{n-1},v_1,w_1$. The
resulting $z'':=(u_i'',v_i'',w_i'')_{i=1,\ldots,n-1}$ can be best
expressed by a matrix vector multiplication $Mz$. The scheme
differs from the one in \cite{Zarif}, where  the PDE is extended
up to and including the vertex from the left, which works well to
discretize the reflectionsless solutions \reff{psia} in case of
\reff{ballist}, but it introduces an asymmetry between the bonds not
present in \pnull.

To integrate the resulting ODEs $\pa_t z=\ri(Mz+\beta |z|^2 z)$,
where $\beta=(\beta_u,\beta_v,\beta_w)$ with obvious meaning, we
use an explicit scheme with stepsize $\tau$ in $t$, namely
\huga{\label{als}
z^{n+1}=z^{n-1}+2\tau(Mz^n+\beta|\tilde{z}^n|^2
z^n),
}
where $\tilde{u}_i=\frac 1 2 (u_{i-1}+u_{i+1})$ and
similar for $\tilde{v}_i$ and $\tilde{w}_i$. For $\tau\le \del^2/4$
this conserves $N(t)$ with high accuracy, and also $H(t)$.

\begin{figure}[t!]
\begin{tabular}{ll}
\ig[width=40mm,height=40mm]{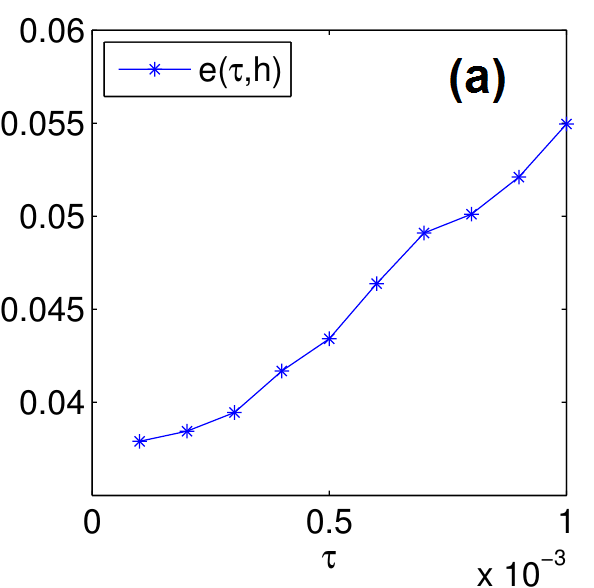}&\ig[width=40mm,height=40mm]{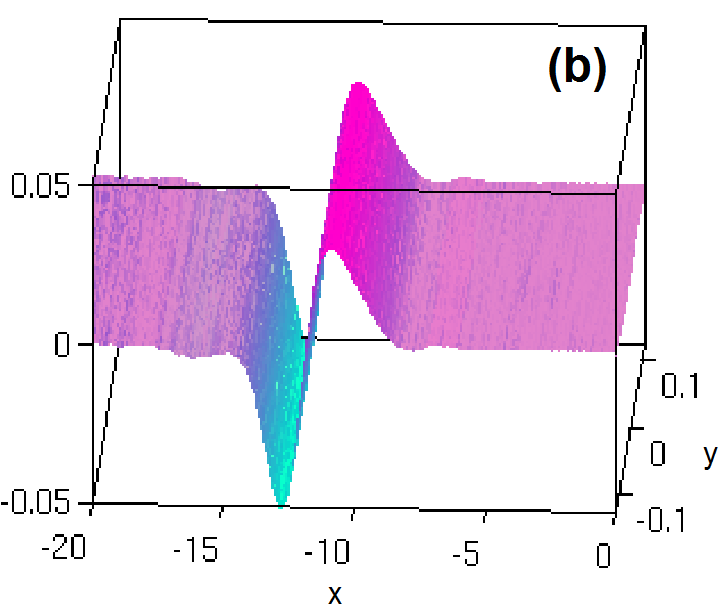}
\end{tabular}
\caption{{Numerical error of the scheme \ref{tint4} over a
straight bond.(a) error dependence on $\tau$; (b)
$|\tilde{\phi}(1,\cdot)|{-}|\phi(1,\cdot)|$,
$\tau{=}10^{-4}$.
\label{f11}}}
\end{figure}

To simulate \peps\ we write it as a 2-component real system for
$z=(u,v)$ where $\psi=u+\ri v$. We set up and discretize the
domain $\Om_\eps$ using routines from {\tt pde2path} \cite{p2p}
which are based on the FEM from the Matlab {\tt pdetoolbox}. For
efficiency it is quite useful to apply some local mesh refinement
near the vertex. We typically work with meshes of 5000-20000
triangles, requiring a maximal mesh size of $\eps/6$ before local
refinement. Eq.~\reff{tbmnls} then translates into the system of
ODEs \huga{\label{dnls} Mz_t=Kz+F(z) } where  $M$ is the mass
matrix, $K=K_{\ri\Del}$ is the stiffness matrix, and $F(z)$ is the
FEM nonlinearity. For the time integration of \reff{dnls} we use a
semilinear trapez rule, i.e., setting $z_n=z(\cdot,t_n),
t_n=n\tau$, with typically $\tau\approx 10^{-4}$ to $10^{-3}$,
\huga{\label{tint4} [M-\frac \tau 2 K]z^{n+1}=[M+\frac \tau 2
K]z^n +\tau F(z^n). } Over relevant time-scales \reff{tint4}
conserves (the discretized version of) $N_\eps$ from
\reff{nepsdef} reasonably well, see, e.g., Fig.~\ref{f2e},
\ref{f2}(a2), and \ref{f21}(b1), (c1), but, as already indicated,
dependent on the discretizations there are some slightly more
significant errors in the numerical soliton speed. To quantify
this, we used \reff{tint4} to propagate a soliton of amplitude and
speed $(\eta,c)=(1,10)$ on a straight bond $x_1\in (-30,0)$ from
$(t,x_{1,s})=(0,-20)$ to position $x_{1,e}\approx -10$ at $t=1$
for various time steps $\tau$ and mesh-sizes $h$, and calculated
the error
$$
e(\tau,h):=\max_x|\, |\tilde{\phi}(1,x)|-|\phi(1,x)|\, |,
$$
where $\tilde{\phi}$ and $\phi$ denote the numerical and the exact
solution, respectively. In reasonable $\tau, h$ ranges $e(\tau,h)$
turns out to be roughly a linear function of $\tau, h$. See
Fig.~\ref{f11}(a) for $e(\cdot,0.04)$ as an example, while (b)
shows $|\tilde{\phi}(1,x)|-|\phi(1,x)|$ for $\tau=10^{-4}$ and
thus illustrates that \reff{tint4} propagates the soliton too
quickly.

 We also tried the relaxation
scheme from \cite{besse04} which conserves $N_\eps$ slightly
better, but becomes computationally much slower, mainly since one
can no longer $LU$-pre-factorize $M-\frac \tau 2 K$. On the other
hand, the stability requirements for explicit schemes like
\reff{als} become prohibitive for fine meshes near the vertex. For
\reff{tint4}, typical computation times for the propagation of a
solitary wave through the network are on the order of 1 minute.

\end{document}